\documentclass[nofootinbib,twocolumn,nofootinbib,showpacs,prd, superscriptaddress]{revtex4}

\usepackage{amsmath}
\usepackage{amssymb}
\usepackage{graphicx}
\usepackage{epstopdf}

\usepackage{inputenc}
\usepackage[usenames,dvipsnames]{color}
\usepackage{hyperref}

\begin{document}


\title{Maximum turnaround radius in $f(R)$ gravity}


\author{Salvatore Capozziello}
\email{capozzie@na.infn.it}
\affiliation{Dipartimento di Fisica, Universit\`a di Napoli  ``Federico II'', Via Cinthia, I-80126, Napoli, Italy.}
\affiliation{Istituto Nazionale di Fisica Nucleare (INFN), Sez. di Napoli, Via Cinthia 9, I-80126 Napoli, Italy.}
\affiliation{Gran Sasso Science Institute (INFN), Viale F. Crispi, 7, I-67100, L'Aquila, Italy.}
\affiliation{Lepage Reseatch Institute, 17. Novembra 1, 08116 Pre\v sov, Slovakia.}

\author{Konstantinos F. Dialektopoulos}
\email{dialektopoulos@na.infn.it}
\affiliation{Dipartimento di Fisica, Universit\`a di Napoli  ``Federico II'', Via Cinthia, I-80126, Napoli, Italy.}
\affiliation{Istituto Nazionale di Fisica Nucleare (INFN), Sez. di Napoli, Via Cinthia 9, I-80126 Napoli, Italy.}
\affiliation{Lepage Reseatch Institute, 17. Novembra 1, 08116 Pre\v sov, Slovakia.}

\author{Orlando Luongo}
\email{luongo@lnf.infn.it}
\affiliation{Istituto Nazionale di Fisica Nucleare (INFN), Laboratori Nazionali di Frascati, Via E. Fermi 40, 00044 Frascati, Italy.}
\affiliation{School of Science and Technology, University of Camerino, I-62032, Camerino, Italy.}


\begin{abstract}
The accelerating behavior of cosmic fluid opposes to the gravitational attraction, at present epoch, whereas standard gravity is dominant at small scales. As a consequence, there exists a \emph{point} where   the effects are counterbalanced, dubbed \emph{turnaround radius}, $r_{\text{ta}}$. By construction, it provides a bound on maximum structure sizes of the observed universe. Once an upper bound on $r_{\text{ta}}$ is provided, i.e. $R_{\text{TA,max}}$, one can check whether cosmological models guarantee structure formation. Here, we focus on $f(R)$ gravity, without imposing \emph{a priori} the form of $f(R)$. We thus  provide an analytic expression for the turnaround radius in the framework of $f(R)$ models. To figure this out, we compute the turnaround radius in two distinct cases: 1) under the hypothesis of static and spherically symmetric space-time, and 2) by using the cosmological perturbation theory. We thus find a criterion to enable large scale structures to be stable in $f(R)$ models, circumscribing the class of $f(R)$ theories as suitable alternative to dark energy. In particular, we get that for constant curvature, the viability condition becomes $R_{\text{dS}}f'(R_{\text{dS}}) \leq 5.48 \Lambda \Rightarrow f'(R_{\text{dS}}) \leq 1.37$, with $\Lambda$ and $R_{\text{dS}}$ respectively the observed cosmological constant and the Ricci curvature. This prescription rules out  models which do not pass the aforementioned $R_{\text{TA,max}}$ limit.
\end{abstract}



\maketitle

\section{Introduction}

The present cosmological picture, the so-called {\it concordance model}, shows  that the  universe  is fueled by some \emph{unknown} fluid, which drives the cosmic dynamics at different epochs \cite{0a}. In particular, the accelerated expansion is today described by some sort of \emph{dark energy}, whose standard origin is presumed to be the cosmological constant, $\Lambda$ \cite{starx1,starx2}. Although appealing and straightforward, the concordance model, dubbed the $\Lambda$CDM, fails to be predictive at quantum regime, i.e. as standard gravity breaks down \cite{rept}. In turn, the fact that at quantum scales gravity cannot be  described by general relativity (GR) suggests more complicated time-dependent dark energy contributions, different from the $\Lambda$CDM model.

\noindent Hence, the concordance paradigm may represent only a first effective explanation of present universe speed up \cite{reviy,rax1,costante1,costante2}. Moreover, structure formation and small perturbations require  non-baryonic dark matter contributions, which dominate over the luminous matter throughout the universe expansion history. To assess the issues of dark energy and dark matter, one wonders whether the net fluid responsible for them is due to geometric corrections, in particular curvature or torsion corrections \cite{francaviglia, raxx,lobo1,lobo2,lobo3, cai}.  In this picture additional geometric terms drive the universe dynamics both at ultra-violet and infra-red scales \cite{starx}. In the case of $f(R)$ gravity,  as curvature corrections propagate, one can imagine that dark energy and dark matter are nothing else but effects of geometry at infrared scales \cite{revvia}. This prescription generalizes GR without introducing barotropic fluids \emph{by hand}, proposing a way to interpret the \emph{dark side of the universe} by a geometrical point of view. In the teleparallel equivalent representation of GR, the so called TEGR, the generalization is achieved considering  functions of the torsion scalar $T$, and then one deals with the so called $f(T)$ gravity \cite{cai}.

Assuming the $f(R)$ gravity, if on the one hand, curvature effects are sub-dominant terms that at certain point lead to a transition at which the expansion starts, on the other hand, a curvature radius related to the scale at which structures form can be defined for $f(R)$ models. In such a way, understanding how curvature corrections affect the  scale of structure formation   is essential to confront dynamics with a given  background\footnote{At small redshifts, dark energy models degenerate to $\Lambda$CDM, so that   higher redshift regimes are essential to discriminate among   viable extensions of  concordance model.}. For these reasons, one invokes the existence of a maximum turnaround radius, say $R_{\text{TA,max}}$, which plays the aforementioned role. The turnaround radius van be defined as the \emph{exact point} at which the attraction of gravity and the repulsion of dark energy cancel each other. For example, considering spherical structures with mass $M$, in the concordance paradigm, the maximum turnaround radius is: $R_{\text{TA,max}} = (3G_NM/\Lambda c^2)^{1/3}$ \cite{Pavlidou:2013zha}. It denotes the maximum size that any  spherically-symmetric structures can assume  in the framework of   $\Lambda$CDM model. A test particle outside this bound follows the Hubble flow, while inside it falls into the gravitational well. As a consequence, the direct comparison of $R_{\text{TA,max}}$ with the present size of cosmic structures enables to test different cosmological models, showing which models are incompatible with structure formation. In  the last few years, this approach has been extensively adopted  to constrain dark energy models and alternative theories of gravity. In particular  Brans-Dicke theory, cubic galileon model, phantom braneworld approach, quintessence, and so forth have been tested against turnaround radius (see for example e.g.  \cite{Pavlidou:2014aia,Faraoni:2015zqa,Faraoni:2015saa,Bhattacharya:2015chc,Lee:2016bec,Tanoglidis:2014lea,Tanoglidis:2016lrj,Bhattacharya:2017ydy,Bhattacharya:2015iha,Velez:2016shi,Bhattacharya:2017yix}). A covariant definition of the maximum turnaround radius, together with the turnaround equation in metric theories of gravity, can be found in \cite{Bhattacharya:2016vur}.

In this paper, we discuss  the turnaround radius in the framework of $f(R)$ gravity. In this perspective, we assume $f(R)$ models which, at certain point, assume a constant curvature, $R= R_{\text{dS}}$. We thus compute the $R_{\text{TA,max}}$ in two different pictures. In the  first approach, we assume  spherically symmetric space-times. In the second, we discuss  the  cosmological case considering perturbations. As expected,  curvature terms account for repulsive effects, mimicking dark energy even at the level of turnaround radius. In this way, we are able to put constraints on generic $f(R)$ models. We find that the observed bounds on $R_{\text{TA,max}}$ are analogous if one considers spherically symmetric space-times or cosmological perturbations\footnote{Throughout the paper we use the Lorentzian signature $(-,+,+,+)$ for the metric.  $G_N$ is the Newton gravitational constant and we set $c=1$ unless otherwise specified.}. We therefore derive limits over the form of $f(R)$ which should guarantee structure formation for any model even in the approximation of quasi-constant curvature.

The paper is organized as follows. In Sec. \ref{sezionedue}, we introduce the role of turnaround radius in cosmology, highlighting its physical properties and the main consequences it may have in structure formation. In Sec. \ref{sezionecinque}, we summarize $f(R)$ gravity with particular attention to cosmological implications. In addition, we evaluate the maximum bound of the turnaround radius in two distinct cases. In the first one, we consider a static and spherically symmetric space-time and we find a bound for the ratio $G_{\text{eff}}/R_{\text{dS}}$, where $G_{\text{eff}}$ is the effective gravitational coupling. In the second case, we take into account cosmological perturbations and we find a constraint for  viable  $f(R)$models. Both the results agree with each other, as well as with  results in literature. Conclusions and perspectives  are  reported in Sec. \ref{sezionesette}.


\section{The  turnaround radius}
\label{sezionedue}

In the concordance $\Lambda$CDM model, the cosmological constant is associated to the vacuum energy and plays a significant role both  at large and small scales. Indeed, in the $\Lambda$-dominated universe, the normalized dark energy abundance is: $\Omega _{\Lambda,0} \simeq 0.73$, affecting structure formation by repulsive interaction\footnote{For alternative effects of repulsive gravity which emerge \emph{without} the need of \emph{any} dark energy contribution see \cite{ioehernprd,ioehernfrw}.}. In this picture, there exists a point, or more precisely a surface, where the attractive force of gravity cancels out with  repulsive effects. This point is  the \emph{turnaround radius}.
A test particle inside it falls into the gravitational well, while outside the test particle follows the Hubble flow. Clearly, the  point of  turnaround radius has an unstable equilibrium.

Alternatively, we can define the turnaround radius as \emph{the scale at which, initially-expanding and gravitationally-bound structures halt their expansion, turn around, and collapse}. This prerogative is a property of structure formation and  its evolution. Thus, it does not depend on  the adopted metric and  gravitational theory. On the contrary, the maximum value of the turnaround radius for a structure of given mass $M$ turns out to be independent of the  cosmic epoch. So that, it is a bound to the maximum size that a structure may have.

In the rest of this section, we sum up the turnaround derivation following the recipe presented in   \cite{Bhattacharya:2016vur}. We first calculate the turnaround radius in the $\Lambda$CDM model using  static coordinates; afterwards, we generalize it to arbitrary theories with static solutions. Furthermore, we compute it using the McVittie metric and, finally, we combine the two results, i.e. static and McVittie, using scalar cosmological perturbations.

Let us consider a spherical mass $M$ in a universe with a cosmological constant, i.e. the Schwarzschild-de Sitter (SdS) space-time
\begin{equation}\label{stsphmetric}
ds^2 = -A(r) dt^2 + B^{-1}(r)dr^2 + r^2 d\Omega ^2 \,,
\end{equation}
with:
\begin{equation}\label{SdSmetric}
B(r) = A(r) = 1-\frac{2 G_N M}{r c^2}- \frac{\Lambda r^2}{3}\,,
\end{equation}
and $d \Omega ^2 = d \theta^2 + \sin ^2 \theta d \phi ^2$. The 4-velocity of a stationary observer is given by $u^{\mu} = \left(A(r)^{-1/2},0,0,0 \right)\,.$
The turnaround radius is defined as the point where, the 4-acceleration of the observer vanishes, i.e.
\begin{equation}\label{tarcond}
R_{\text{TA,max}} = \left(\frac{3G_N M}{\Lambda c^2} \right)^{1/3}\,.
\end{equation}

Since the SdS-metric is not necessarily a solution of any theory of gravity, this result does not hold in general. In theories that obey the weak equivalence principle and have solutions of the form \eqref{stsphmetric}, the turnaround radius, following the same rationale as before, i.e. the vanishing 4-acceleration of a stationary observer, can be computed from
\begin{equation}\label{tar}
A'(r)|_{r = R_{\text{TA,max}}} = 0\,,
\end{equation}
which is an algebraic equation in terms of $r$. Note that, even if the solutions are not static, they can always be diagonalized in the form \eqref{stsphmetric} and thus the formula \eqref{tar} is still valid. In that case, however, the turnaround radius becomes time-dependent.

Eq. \eqref{tar} only holds for this specific coordinate system. Thus, it is more convenient if we find a covariant version of it. So that, for a stationary observer in a spherically symmetric space-time, the point at which $u^{\mu}\nabla_{\mu}u^{\nu} = 0$ is identified with the turnaround radius.

\noindent Starting from the fact that we want to extend the  result to  cosmology, we consider a McVittie space-time, which describes a spherical mass $M$ in an expanding universe. Its line element is given by
\begin{equation}\label{McVittie}
ds^2 =-\left( \frac{1-\mu}{1+\mu}\right)^2 dt^2 + (1+\mu)^4 a^2 \left(dr^2 + r^2 d \Omega ^2 \right)\,,
\end{equation}
where $\mu = G_N M/(2 a r)$. If we consider a pressureless fluid with $8 \pi G_N \rho = \Lambda$,  adopting the coordinate transformations
\begin{subequations}
\begin{align}\label{coordtrans1}
\tilde{t}(t,r) &= t + Q(\tilde{r})\,, \\
\tilde{r} (t,r) &= \left(1+\mu\right)^2ar \,, \label{coordtrans2}
\end{align}
\end{subequations}
with $Q(\tilde{r})$ satisfying the condition
\begin{equation}
\frac{\partial Q}{\partial \tilde{r}} = \frac{\sqrt{\Lambda /3} \tilde{r}}{\left( 1- \frac{2 G_N M}{\tilde{r}}- \frac{\Lambda }{3}\tilde{r}^2\right) \sqrt{1- \frac{2 G_N M }{\tilde{r}}}}\,,
\end{equation}
 we recover the SdS-metric given by \eqref{stsphmetric} and \eqref{SdSmetric}.

In order to see how the turnaround radius looks like in this space-time, we  need to transform   results in SdS-space-time adopting the inverse transformations \eqref{coordtrans1},\eqref{coordtrans2}. The observer  4-velocity  becomes now
\begin{equation}\label{4velocity}
u^{\mu} = \frac{1+\mu}{\sqrt{\left(1-\mu \right)^2 - H^2\left(1-\mu \right)^6a^2r^2}} \left( 1,-rH,0,0\right)\,,
\end{equation}
and using $u^{\mu}\nabla_{\mu}u^{\nu} = 0$ we find the turnaround radius in McVittie space-time
\begin{equation}
2 \mu = (1+ \mu)^6 H^2 a^2 r^2\,,
\end{equation}
which takes the form \eqref{tarcond} if we use \eqref{coordtrans1},\eqref{coordtrans2}.

In general,  the SdS-metric and  static, spherically symmetric metrics that can be transformed into McVittie metric, cannot be solutions in arbitrary modified theories of gravity. Nonetheless, a perturbed Friedmann-Robertson-Walker (FRW) can be used for characterizing small perturbations.

Indeed, the McVittie metric, for $\mu \ll 1$, can be seen as a perturbed FRW metric. Identifying $\Phi = \Psi = - 2 \mu = - G_N M/(ar)$, Eq. \eqref{McVittie} can be expressed as the perturbed FRW metric in the Newtonian gauge
\begin{equation}\label{FRWpert}
ds^2 = - \left(1+2 \Phi(t,r^i) \right) dt^2 + a^2 \left(1- 2 \Psi(t,r^i) \right) \gamma_{ij}dx^i dx^j\,.
\end{equation}
In Cartesian coordinates $\gamma_{ij}dx^idx^j = dx^2 + dy^2 +dz^2$ and $r^i = \{ x,y,z\}$,  assuming spherical symmetry, it is  $R = r^i = \left(x^2+y^2+z^2\right)^{1/2}$. In spherical coordinates, it is  $r^i = r$ and $\gamma_{ij}dx^i dx^j = dr^2 + r^2 d\Omega^2$.

Taking advantage of this fact, we find\footnote{These assumptions impose the region of interest, which is far beyond both horizons. We assume a structure whose maximum size (maximum turnaround radius) is much larger than its Schwarzschild horizon and much smaller than the cosmological horizon.} the 4-velocity \eqref{4velocity} in the limit $\mu \ll 1$ and $aHr \ll 1$. Using the geodesic equation $u^{\mu}\nabla_{\mu}u^{\nu} = 0$, we finally get the turnaround for spherically symmetric space-times in cosmological framework, that is
\begin{equation}\label{cosmtar}
a^2 \left(H^2 + \dot{H}\right) r = \frac{\partial \Phi}{\partial r}\,.
\end{equation}
Specifically, given a Hubble parameter and a potential $\Phi$, Eq. \eqref{cosmtar} gives the maximum turnaround radius $R_{\text{TA,max}} = ar$. For the general turnaround equation, including non-spherical space-times,  we refer to \cite{Bhattacharya:2016vur}.

A quick look at the previous results shows that the two outcomes, i.e. Eq. \eqref{cosmtar} and \eqref{tarcond}, are equivalent. In fact, in the $\Lambda$CDM model, the potential $\Phi$ is given by $\Phi = - G_N M/(ar)$ and the Hubble constant by $H = \sqrt{\Lambda/3}$. 

Before  proceeding with our considerations,  let us see how  the turnaround radius can be used to constrain different cosmological models. To do so, we can compare the masses and the radii of known cosmic structures. Observationally the turnaround radius is the zero-velocity surface of a structure and can be determined by observations of the Hubble flow in structures that consist of, at least, several galaxies. The concept of mass needs some more investigation. The usual procedure for the mass estimation  of a structure is the combination of turnaround radius measurements and predictions of the spherical collapse models in the given cosmology. The straightforward  application of this approach can be very complicated. Alternatives are provided by summing up the masses of the constituent galaxies, strong or weak gravitational lensing studies, or X-ray spectroscopy studies. Even a single observation of a structure that exceeds the theoretically predicted bound (assuming that the confidence of the measurement is really high) would be enough to constrain a cosmological model.

In the $\Lambda$CDM model, the obtained formula for the turnaround radius \eqref{tarcond}, even for structures as big as $10^{15}M_{\odot}$ (Virgo cluster), is $10\%$ greater than its actual observed size. A detailed analysis about the structures used as well as relevant plots can be found in \cite{Pavlidou:2013zha}. This means that the turnaround radius in any theory can be at most $10\%$ smaller than in  $\Lambda$CDM model. The observed size of  structures  is, in general,  at low redshift, i.e. when the structures have almost reached their maximum size. This means that  we do not need data from higher redshifts or from the  early stages of the universe to evaluate the turnaround radius.


\section{Turnaround radius in $f(R)$ gravity}
\label{sezionecinque}

We have reviewed the definition of the turnaround radius. In particular, we proved that Eq. \eqref{cosmtar} can give the maximum turnaround radius in any theory of gravity where the Einstein Equivalence Principle holds. In this section, we compute the turnaround radius in $f(R)$ gravity taking into account  two cases. First we consider a static and spherically symmetric space-time,  after, we investigate the turnaround radius in the context of cosmological perturbations.

Briefly,  $f(R)$ gravity is obtained by substituting  the Hilbert-Einstein  action, linear in the Ricci scalar $R$,  with an arbitrary function of $R$, i.e.
\begin{equation}\label{fRaction}
\mathcal{S} = \int d^4 x\sqrt{-g}\left(\frac{1}{16 \pi G_N}f(R) + \mathcal{L}_m \right) \,.
\end{equation}
By varying the above action with respect to the metric, we get the field equations 
\begin{align}\label{eqmot}
f'(R) R_{\mu\nu} - \frac{1}{2}f(R)g_{\mu\nu}+ \Big(g_{\mu\nu}\square - &\nabla_{\mu}\nabla_{\nu} \Big)f'(R) =\nonumber \\
&= 8 \pi G_N T_{\mu\nu}^M\,,
\end{align}
where $T_{\mu\nu}^M$ is the energy-momentum tensor of matter. Another interesting equation, is  the trace of  Eq. \eqref{eqmot},
\begin{equation}\label{treq}
f'(R)R-2 f(R)+3\square f'(R) = 8\pi G_N T^M\,,
\end{equation}
which relates the Ricci scalar to $f(R)$ and its derivative in $R$, that is $f'(R)$.

Several efforts have been spent to get constraints on viable $f(R)$ models (see for example  \cite{starx,rept,Clifton:2011jh,repva}). In particular, we can  assume $f''(R)>0$, in order to avoid tachyonic instabilities, and $f'(R)>0$, to make the theory ghost-free.

\subsection{The spherically symmetric case}
\label{sphersymsec}

As  mentioned in the Introduction, we can restrict our attention to $f(R)$ models   with constant curvature solutions, i.e. $R = R_{\text{dS}} = const$. This $R_{\text{dS}}$ is the solution of the trace Eq. \eqref{treq}, which, for constant curvature, takes the form
\begin{equation}\label{treq2}
Rf'(R) - 2 f(R) = 0\,.
\end{equation}
Solutions\footnote{Such solutions  exist in the constant curvature $f(R)$ models. They imply the validity of the  Birkhoff theorem that does not always hold in $f(R)$ gravity  \cite{Nzioki:2013lca,Capozziello:2011wg}.}  to this equation are known as de Sitter points \cite{DeFelice:2010aj,Cognola:2005de}. Moreover, following \cite{Motohashi:2011wy,Bamba:2010iy}, nonlinear models should have stable de Sitter points at late times. From the Eq. \eqref{eqmot}, using Eq. \eqref{treq2}, we get
\begin{equation}
R_{\mu\nu} = \frac{f(R_{\text{dS}})}{2 f'(R_{\text{dS}})}g_{\mu\nu} = \frac{R_{\text{dS}}}{4}g_{\mu\nu}\,,
\end{equation}
which gives rise to an effective cosmological constant of the form
\begin{equation}
\Lambda_{eff} = \frac{f(R_{\text{dS}})}{2f'(R_{\text{dS}})} = \frac{R_{\text{dS}}}{4}\,.
\end{equation}
The stability of these points is discussed in \cite{DeFelice:2010aj,Amendola:2006we}. They asymptotically  approach the (anti)-de Sitter space with $\Lambda = \Lambda_{\text{eff}}$.

If we consider a static and spherically symmetric  metric of the form \eqref{stsphmetric}, the equations of motion \eqref{eqmot}, for both the $(0,0)$ and $(1,1)$ components, together with the trace equation, are respectively
\begin{subequations}
\begin{align}\label{eq1}
\Big(\frac{1}{2}Af(R) + \frac{A'f'(R)}{rB}&-\frac{A'^2f'(R)}{4 AB}-\frac{A'B'f'(R)}{4B^2}+\nonumber \\
&+\frac{f'(R)A''}{2B}\Big)\bigg\vert _{R=R_{\text{dS}}} = 0 \,, \\ \label{eq2}
\Big( - \frac{1}{2}Bf(R) + \frac{A'^2 f'(R)}{4A^2}&+\frac{B'f'(R)}{rB}+\frac{A'B'f'(R)}{4AB}-\nonumber \\
&-\frac{f'(R)A''}{2A} \Big)\bigg\vert_{R = R_{\text{dS}}} = 0 \,,\\ \label{eq3}
\Big(R f'(R) - 2 f(R)\Big)&\bigg\vert _{R=R_{\text{dS}}} = 0\,.
\end{align}
\end{subequations}
We can easily see, from  the  first two equations, \eqref{eq1} and \eqref{eq2}, that
\begin{equation}\label{Bsol}
B(r) = \frac{c}{A(r)}\,,
\end{equation}
where $c$ is an integration constant. The constant can  be set equal to 1  to recover the Minkowski space-time as limiting case. Hence, Eq.  \eqref{eq2} provides:
\begin{align}\label{SdSfR}
A(r) &= a_1 - \frac{a_2}{r} - \frac{r^2}{6}\frac{f(R)}{f'(R)}\bigg\vert _{R_{\text{dS}}} \nonumber \\
& = a_1 - \frac{a_2}{r} - \frac{R_{\text{dS}}}{12}r^2 \,,
\end{align}
where, in the second line,  we used Eq. \eqref{eq3}, and  $a_1,\,a_2$ are constants. Without losing generality, we choose $a_1 = 1$ and $a_2 = 2 G_{\text{eff}} M/c^2$, where $G_{\text{eff}}$ is the effective gravitational coupling. This allows to recover a Schwarzschild-like solution in the $R_{\text{dS}} \rightarrow 0$ limit.

It is worth-mentioning that in the Hilbert-Einstein  limit, i.e. $f(R) = R - 2 \Lambda$, we can set $R_{\text{dS}} =  4 \Lambda > 0$, since $\Lambda$ is positive-definite. As a consequence, we recover the known Schwarzschild-de Sitter solution \eqref{SdSmetric}.

Finally, the maximum turnaround radius for any $f(R)$ model with $R = R_{\text{dS}}$ is given by Eq. \eqref{tar}
\begin{equation}\label{fRtar}
A'(R_{\text{TA,max}}) = 0 \Rightarrow R_{\text{TA,max}} = \left( \frac{12 G_{\text{eff}} M}{R_{\text{dS}}}\right)^{1/3}\,.
\end{equation}
As we already mentioned in Sec. \ref{sezionedue}, the maximum turnaround radius in any alternative theory of gravity can be, at most, $10\%$ smaller than the corresponding one  in GR. Thus, by comparing \eqref{fRtar} with \eqref{tarcond},  we get the following constraint
\begin{equation}\label{constraint2}
\frac{G_{\text{eff}}}{R_{\text{dS}}} \geq \frac{0.18 G_N}{\Lambda}\,.
\end{equation}
At this point, a short comment on Eq. \eqref{treq2} is needed.  Solutions  $R_{\text{dS}} = 0$,  are not excluded \emph{a priori}. However, these are trivial Minkowski solutions, instead of de Sitter ones, leading to neither expanding nor contracting universes. In this case, the maximum turnaround radius cannot be defined. In the next section, we will see that scalar cosmological perturbations give the same result as Eq. \eqref{constraint2}, together with a specific form for the gravitational coupling.


\subsection{The cosmological case}

In the previous section, we studied the turnaround radius derived from static and spherically symmetric space-times in $f(R)$ gravity. However, we want to find a more general formula for the turnaround radius and thus we turn to cosmology. Specifically, in this section, we are going to use Eq. \eqref{cosmtar} in order to see, whether we can extend  the result \eqref{constraint2}.

To this end, let us  consider a spherical cosmic structure described by a perfect fluid with non-relativistic matter, i.e. with pressure $P = 0$, and all its mass is assumed to be at the center, $r = 0$. We perturb this structure by a test fluid and we study its dynamics. The whole configuration can be described by a perturbed FRW metric, which in conformal Newtonian gauge, can be expressed in the form \eqref{FRWpert}. For a detailed discussion about the relation between static and comoving coordinates, as well as the relation of the metric \eqref{FRWpert}, with the Bardeen potentials, see \cite{Velez:2016shi}.

The  homogeneous background Eqs. \eqref{eqmot} in a  FRW flat space-time $g_{\mu\nu} = \text{diag}(-1,a^2(t),a^2(t),a^2(t))$, and with non-relativistic matter, i.e. $T_{\mu\nu} = \text{diag}(-\rho,0,0,0)$, are given by
\begin{subequations}
\begin{align}\label{back1}
3 f' H^2 &= 8 \pi \rho + \frac{1}{2}(Rf' - f) - 3 f'' H \dot{R} \,, \\ \label{back2}
f' \left(2 \dot{H}+ 3H^2\right) &= \frac{1}{2}(Rf' - f) - f''' \dot{R}^2- f'' \ddot{R}
\end{align}
\end{subequations}
together with the continuity equation
\begin{equation}\label{back3}
\dot{\rho} + 3 H \rho = 0\,.
\end{equation}
The prime denotes differentiation with respect to $R$, the dot with respect to time $t$, and $H = \dot{a}/a$ is the Hubble parameter. In this case, the Ricci scalar is
\begin{equation}\label{back4}
R = 6 \left(2H^2+\dot{H}\right)\,,
\end{equation}
and Eq. \eqref{back3} gives $\rho = \rho_0/a^3$, where $\rho_0$ is the constant rest-mass density.

The perturbed energy momentum tensor is given by $T_{00} = -\rho-\delta \rho$, $T_{0i} = -\rho \delta \upsilon_i$ and $T_{ij} = 0$. Thus, the perturbed equations are \cite{Hwang:2001qk,Eingorn:2014laa,DeFelice:2010aj}
\begin{widetext}
\begin{subequations}
\begin{gather}\label{pert1}
-\frac{\Delta \Psi}{a^2}+3H\left(H\Phi + \dot{\Psi}\right) = -\frac{1}{2f'}\left[ 8\pi G_N\delta \rho + \left(3H^2+3\dot{H}+\frac{\Delta}{a^2}\right)\delta f' -3H\left(\dot{\delta f'}- 2\dot{f'}\Phi\right) +3 \dot{f'}\dot{\Psi}  \right]\,,\\ \label{pert2}
H\Phi + \dot{\Psi} = \frac{1}{2f'}\left[8\pi G_N \rho \delta \upsilon + \dot{\delta f'}-H\delta f'-\dot{f'}\Phi  \right]\,,\\ \label{pert3}
\Psi - \Phi = \frac{\delta f'}{f'}\,,\\
3\left(\dot{H}\Phi+H\dot{\Phi}+\ddot{\Psi}\right)+6H\left(H\Phi+\dot{\Psi}\right)+\left(3H+\frac{\Delta}{a^2}\right)\Phi = \frac{1}{2f'}\Big[8 \pi G_N \delta \rho + 3\ddot{\delta f'}+3H\dot{\delta f'}-\left(6H^2+\frac{\Delta}{a^2}\right)\delta f'-\nonumber \\ \label{pert4}
-3\dot{f'}\left(\dot{\Phi}+2H\Phi+\dot{\Psi}\right)-6\ddot{f'}\Phi \Big]\,,\\ \label{pert5}
\ddot{\delta f'}+3H\dot{\delta f'}-\left( \frac{\Delta}{a^2}+\frac{R}{3}\right)\delta f' = \frac{8\pi G_N}{3}\delta \rho+\dot{f'}\left(\dot{\Phi}+6H\Phi+3\dot{\Psi} \right) + 2\ddot{f'}\Phi -\frac{1}{3}f' \delta R\,,
\end{gather}
\end{subequations}
where \begin{equation} \label{pert8}
\delta R = -2 \left[3 \left(\dot{H}\Phi+H\dot{\Phi}+\ddot{\Psi} \right)+12 H\left(H\Phi+\dot{\Psi} \right)+\frac{\Delta}{a^2}\left( \Phi - 2\Psi\right)+3\dot{H}\Phi \right]\,.
\end{equation}
\end{widetext}
As before, $f' = df/dR$, dot denotes differentiation with respect to cosmic time $t$ and $\Delta$ is the Laplacian in comoving coordinates.

We can safely use the quasi-static approximation \cite{Tsujikawa:2007tg,Tsujikawa:2007gd} according to which the inhomogeneities $\Phi$ and $\Psi$ are primarily produced by the spatial distribution of matter. This means that the spatial derivatives of the fields are dominant in the equations (see \cite{DeFelice:2010aj}). In this way,  Eq. \eqref{pert1}, \eqref{pert4}, \eqref{pert5} yield\footnote{Eq. \eqref{pert2} means that, in the considered order of approximation, the peculiar velocities of the test fluid are ignored and Eq. \eqref{pert3} is in the correct order and it is satisfied, \textit{a posteriori}, by Eq. \eqref{PhiPsi}.}
\begin{subequations}
\begin{gather}\label{quazi1}
\frac{\Delta \Psi}{a^2} =  \frac{1}{2f'}\left[8 \pi G_N \delta \rho+\frac{\Delta \delta f'}{a^2} \right]\,, \\ \label{quazi2}
\frac{\Delta \Phi}{a^2} = \frac{1}{2f'}\left[8\pi G_N \delta \rho - \frac{\Delta \delta f'}{a^2} \right]\,,\\ \label{quazi3}
-\frac{\Delta \delta f'}{a^2}- \frac{R\delta f'}{3} = \frac{8\pi G_N \delta \rho}{3} - \frac{1}{3}f' \delta R\,,
\end{gather}
\end{subequations}
where $\delta R = -\frac{2\Delta}{a^2} \left(\Phi-2\Psi \right)$. By using $\delta f' = f'' \delta R$, \eqref{quazi1} and \eqref{quazi2} immediately give
\begin{align}\label{PhiPsi}
\Psi = \frac{f''}{2f'} \delta R + \phi\,,\quad \Phi = -\frac{f''}{2f'} \delta R + \phi\,,
\end{align}
where $\phi$ satisfies the Poisson-like equation
\begin{equation}\label{Poissonlike}
\Delta \phi = \frac{4 \pi G_N a^2}{f'} \delta \rho\,.
\end{equation}
In addition, \eqref{quazi3} gives
\begin{equation}\label{Helmholtz}
\left( \Delta - a^2 M^2\right)\delta R = - \frac{8\pi G_N a^2}{3f''}\delta \rho\,,
\end{equation}
where we defined the mass term (of the scalar degree of freedom) as
\begin{equation}\label{massR}
M^2 = \frac{1}{3}\left[\frac{f'(R_{\text{dS}})}{f''(R_{\text{dS}})}-R_{\text{dS}} \right] \,.
\end{equation}
We notice that Eq.\eqref{Helmholtz} is a modified-Helmholtz equation and, if we set $\delta \rho (r) \sim \mathcal{M}\delta(r)/a^3$, where $\mathcal{M}$ is the mass of the structure, Eq. \eqref{Helmholtz} and Eq. \eqref{Poissonlike} give respectively
\begin{equation}
\delta R = \frac{2 G_N \mathcal{M} }{3f''a r}e^{-aMr}\,,\quad \phi = - \frac{G_N \mathcal{M}}{f' a r} \,,
\end{equation}
where we assumed that $\delta R \rightarrow 0$ when $r$ is large. Finally, the gravitational potentials from \eqref{PhiPsi} become
\begin{equation}\label{Phifinal}
\Phi = -\frac{G_N\mathcal{M}}{f'ar}\left(1+\frac{e^{-aMr}}{3}\right)\,,\, \Psi= -\frac{G_N\mathcal{M}}{f'ar}\left(1-\frac{e^{-aMr}}{3}\right)\,.
\end{equation}
Before proceeding to the calculation of the turnaround radius through Eq.\eqref{cosmtar}, let us  comment on these results. First of all, Eq. \eqref{Phifinal} shows that the effective gravitational coupling is
\begin{equation}\label{Geff}
G_{\text{eff}} = \frac{G_N}{f'(R_{\text{dS}})}\left(1+\frac{e^{-M r_{ph}}}{3} \right)\,,
\end{equation}
where $r_{ph} = a r$ is the physical distance. So, even in the weak field limit, deviations from GR, i.e. $f'(R) \neq 1$,  become evident. Apart from this, we see that the inhomogeneities caused by the test fluid, together with the non-linearities of the theory, contribute as a Yukawa-like correction to the gravitational fields. This  has been observed in different contexts \cite{Stelle:1976gc,Mannheim:1988dj}. Such corrections can effectively explain  the flat rotation curves of galaxies, without invoking any exotic form of matter \cite{Capozziello:2012ie}. Moreover, it is worth noticing that, with growing $r$, the Yukawa corrections vanish, while $\Phi$ and $\Psi$ evolve towards a McVittie-like form, which is expected in scalar cosmological perturbations.

Let us proceed now with the  calculation of  the turnaround radius. Eq. \eqref{cosmtar} gives 
\begin{equation}\label{eq4}
r_{ph}^3 = \frac{12 G_N \mathcal{M} }{R_{\text{dS}}f'(R_{\text{dS}})}\left[1+\frac{e^{-Mr_{ph}}}{3}\left( 1+\frac{1}{3}Mr_{ph}\right) \right]\,.
\end{equation}
As already mentioned, $M$ is the effective mass related to the further  degree of freedom of $f(R)$ gravity. Thus,  $Mr \ll 1$ means that the mass of the scalar field is small and the related effective length is very large  with respect to the  the Solar System scale. On the other hand, if the  mass $M$ is large, it cannot have observable effects  at late times\footnote{One can claim that the mass $M$  is a function of the curvature, and then of the energy density, and thus it is  small at cosmological scale and large at Solar System scales like in the  so-called  Chameleon Mechanism \citep{Khoury:2003rn, CapTsu}.}. Hence, Eq. \eqref{eq4} becomes of zeroth order in $Mr$
\begin{gather}\label{fRtar2}
R _{\text{TA,max}} \simeq \left(\frac{12 G_N \mathcal{M}}{R_{\text{dS}}f'(R_{\text{dS}})} \right)^{1/3}\,.
\end{gather}
Clearly,  Eq. \eqref{fRtar2} is the same of Eq. \eqref{fRtar} for $G_{\text{eff}} = G_N/f'(R_{\text{dS}})$, which is \eqref{Geff} for $Mr \ll 1$.
Thus the constraint \eqref{constraint2} becomes
\begin{align}
R_{\text{dS}}f'(R_{\text{dS}}) &\leq 5.48 \Lambda\,.
\end{align}
This is the main result of the paper; the maximum turnaround radius can be used  to set a further  criterion for the viability of $f(R)$ models, through the stability of cosmic structures. As we already mentioned in Sec. \ref{sphersymsec}, for late times, when the matter density becomes negligible compared to $\Lambda$, the  Ricci curvature scalar takes the value $R_{\text{dS}} = 4 \Lambda$; thus the upper bound for the first derivative of the model is $f'(R_{\text{dS}}) \leq 1.37$ (which, of course, in GR is $f'(R_{\text{dS}})=1$). Summarizing, viable  $f(R)$ models are those which obey the following criteria
\begin{equation}\label{fRconst}
0< f'(R)\leq 1.37 \,,\quad f''(R)>0\,\, \text{for} \,\, R \geq R_{\text{dS}} \geq 0\,.
\end{equation}
As an example we can consider a power-law model 
\begin{equation}\label{powerlaw}
f(R) = R+ \alpha R^{n}\,, \quad n>0\,,
\end{equation}
where $n$ is a  positive real number and $\alpha$ is a dimensional  constant. For $n=0$, $\alpha$ plays the role of  cosmological constant. In the literature,  there are several constraints on $n$ (e.g. \cite{Dialektopoulos:2017pgo} and references therein) and thus we will consider only $n>2$. From the constraint \eqref{fRconst}, we find, for example,  that for $n=3$, it is $\alpha \lesssim 1/(20 \Lambda^2)$, as shown in Fig.1.
\begin{figure}
\includegraphics[scale=0.6]{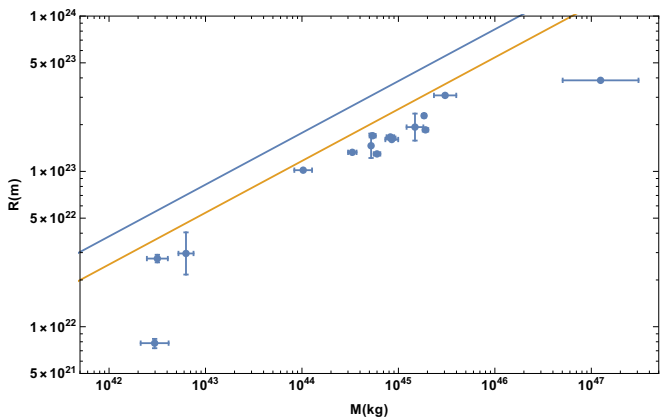}
\caption{In the picture,  the observational values of the turnaround radius for some  astrophysical  structures are reported. Details about the  used data can be found in \cite{Pavlidou:2013zha,Pavlidou:2014aia} and references therein. The theoretical bound of the maximum turnaround radius in $\Lambda$CDM model (blue line), as well as in the power-law $f(R)$model with $n=3$ (orange line) are reported.}
\end{figure}
\\

\section{Final outlooks and perspectives}
\label{sezionesette}
In this work, we considered  $f(R)$ gravity to find out a turnaround radius for cosmological structures. Standard requirements of such theories are that $f''(R)>0$ to avoid tachyonic instabilities and $f'(R)>0$ to let the theory be ghost-free. Several bounds and conditions may be put on the $f(R)$ derivatives to check whether it is possible to circumscribe the functional form of $f(R)$. This is a consequence of the fact that the form of $f(R)$ is not known \emph{a priori} and deserves reconstruction techniques to be determined. Hence, we studied whether it is possible to constrain $f(R)$ invoking a maximum turnaround radius, i.e. a radius that fixes the  stability of large scale structures with respect to the background cosmic evolution. To do so, we considered two approaches: the first concerning a spherically symmetric metric and the second adopting cosmic perturbations.  In both cases, we got analogous outcomes which allow the existence of stable structures according to a stability criterion which is  $f'(R_{\text{dS}}) \leq 1.37$. To achieve this result, we first focused  on $f(R)$ models with  constant curvature solutions. In this case,  the result is analytical. The same result is valid if curvature is approximately constant as in the case of several astrophysical cases. In a forthcoming paper, considering also the study reported in \cite{salzano}, the present criterion will be confronted with   galaxy clusters.

\acknowledgements
S.C. and K.F.D. acknowledge COST action CA15117 (CANTATA), supported by COST (European Cooperation in Science and Technology). They are also partially supported by the INFN sezione di Napoli (TEONGRAV).



\begin{thebibliography}{99}


\bibitem{0a}
J. V. Narlikar, T. Padmanabhan, Ann. Rev. of Astr. and Astroph., {\bf 39}, 211-248, (2001).

\bibitem{starx1}
V. Sahni, A. A. Starobinsky, Int. J. Mod. Phys. D, 9, (2000), 373.

\bibitem{starx2}
T. D. Saini, S. Raychaudhury, V. Sahni, A. A. Starobinsky, Phys. Rev. Lett., 85, (2000), 1162.

\bibitem{rept}
S. Capozziello, M. De Laurentis, Phys. Rept. 509,  (2011) 167.


\bibitem{reviy}
M. Li, X. D. Li, S. Wang, Y. Wang, M. Li, X. D. Li, S. Wang, Y. Wang, Commun. Theor. Phys., 56, (2011), 525.

\bibitem{rax1}
U. Alam, V. Sahni, A. A. Starobinsky, JCAP, 0406, (2004), 008.

\bibitem{costante1}
S. M. Carroll, Liv. Rev. Rel., 4, (2001), 1.

\bibitem{costante2}
S. M. Carroll, Ann. Rev. Astron. Astroph., 30, (1992), 499.

\bibitem{raxx}
A. A. Starobinsky, JETP Lett., 86, (2007), 157.

\bibitem{francaviglia}
S. Capozziello, M. Francaviglia, Gen.Rel.Grav. 40 (2008) 357.

\bibitem{lobo1}
O. Bertolami, C. G. Boehmer, T. Harko, and F. S. N. Lobo, Phys. Rev. D, 75, (2007), 104016.


\bibitem{lobo2}
C. G. Boehmer, T. Harko, and F. S. N. Lobo, JCAP, 0803, (2008), 024.


\bibitem{lobo3}
C. G. Boehmer, L. Hollenstein, and F. S. N. Lobo, Phys. Rev. D, 76, (2007), 084005.

\bibitem{cai}
Y. Cai, S. Capozziello, M. De Laurentis, E. N. Saridakis,  Rept. Prog. Phys. 79 (2016), 106901.

\bibitem{starx}
S. Nojiri, S. D. Odintsov, Phys. Rept. 505, (2011)  55.



\bibitem{revvia}
S. Capozziello, M. De Laurentis, O. Luongo, A. C. Ruggeri, Galaxies, 1(3), (2013), 216.


\bibitem{Pavlidou:2013zha}
  V.~Pavlidou and T.~N.~Tomaras,  JCAP {\bf 1409} (2014) 020.

\bibitem{Pavlidou:2014aia}
  V.~Pavlidou, N.~Tetradis and T.~N.~Tomaras, JCAP {\bf 1405} (2014) 017.


\bibitem{Faraoni:2015zqa}
  V.~Faraoni,
  Phys.\ Dark Univ.\  {\bf 11} (2016) 11

\bibitem{Faraoni:2015saa}
  V.~Faraoni, M.~Lapierre-Léonard and A.~Prain,
  JCAP {\bf 1510} (2015) no.10,  013

\bibitem{Bhattacharya:2015chc}
S.~Bhattacharya, K.~F.~Dialektopoulos, and T.~N.~Tomaras,
JCAP {\bf 1605} (2016) 036


\bibitem{Lee:2016bec}
  J.~Lee and B.~Li,
  Astrophys.\ J.\  {\bf 842} (2017) no.1,  2

\bibitem{Tanoglidis:2014lea}
  D.~Tanoglidis, V.~Pavlidou and T.~Tomaras, JCAP {\bf 1512} (2015) 12,  060.

\bibitem{Tanoglidis:2016lrj}
  D.~Tanoglidis, V.~Pavlidou and T.~Tomaras, arXiv:1601.03740 [astro-ph.CO].


\bibitem{Bhattacharya:2017ydy}
  S.~Bhattacharya and S.~R.~Kousvos,
  Phys.\ Rev.\ D {\bf 96} (2017) no.10,  104006

\bibitem{Bhattacharya:2015iha}
S.~Bhattacharya, K.~F.~Dialektopoulos, A.~E.~Romano, and T.~N.~Tomaras,
Phys.\ Rev.\ Lett.\ {\bf 115} (2015)  181104

\bibitem{Velez:2016shi}
  C.~S.~Velez and A.~E.~Romano,
  arXiv:1611.09223 [gr-qc].

\bibitem{Bhattacharya:2017yix}
  S.~Bhattacharya and T.~N.~Tomaras,
  Eur.\ Phys.\ J.\ C {\bf 77} (2017) no.8,  526

\bibitem{Bhattacharya:2016vur}
  S.~Bhattacharya, K.~F.~Dialektopoulos, A.~E.~Romano, C.~Skordis and T.~N.~Tomaras,
  JCAP {\bf 1707} (2017) no.07,  018

\bibitem{ioehernfrw}
O. Luongo, H. Quevedo, arXiv[gr-qc]:1507.06446 (2015).

\bibitem{ioehernprd}
O. Luongo, H. Quevedo, Phys. Rev. D, {\bf 90}, 8, (2014), 08403.

\bibitem{Clifton:2011jh}
  T.~Clifton, P.~G.~Ferreira, A.~Padilla and C.~Skordis, Phys.\ Rept.\  {\bf 513} (2012) 1.

\bibitem{repva}
S. Nojiri, S. D. Odintsov, V. K. Oikonomou,  Phys. Rept. 692, (2017)  1.

\bibitem{DeFelice:2010aj}
  A.~De Felice and S.~Tsujikawa,
  Living Rev.\ Rel.\  {\bf 13} (2010) 3

\bibitem{Cognola:2005de}
  G.~Cognola, E.~Elizalde, S.~Nojiri, S.~D.~Odintsov and S.~Zerbini,
  JCAP {\bf 0502} (2005) 010

\bibitem{Motohashi:2011wy}
  H.~Motohashi, A.~A.~Starobinsky and J.~Yokoyama,
  JCAP {\bf 1106} (2011) 006

\bibitem{Bamba:2010iy}
  K.~Bamba, C.~Q.~Geng and C.~C.~Lee,
  JCAP {\bf 1011} (2010) 001

\bibitem{Amendola:2006we}
  L.~Amendola, R.~Gannouji, D.~Polarski and S.~Tsujikawa,
  Phys.\ Rev.\ D {\bf 75} (2007) 083504
  doi:10.1103/PhysRevD.75.083504
  [gr-qc/0612180].

\bibitem{Nzioki:2013lca}
  A.~M.~Nzioki, R.~Goswami and P.~K.~S.~Dunsby,
  Phys.\ Rev.\ D {\bf 89} (2014) no.6,  064050

\bibitem{Capozziello:2011wg}
  S.~Capozziello and D.~Saez-Gomez,
  Annalen Phys.\  {\bf 524} (2012) 279

\bibitem{Hwang:2001qk}
  J.~c.~Hwang and H.~r.~Noh,
  Phys.\ Rev.\ D {\bf 65} (2002) 023512
  doi:10.1103/PhysRevD.65.023512
  [astro-ph/0102005].

\bibitem{Eingorn:2014laa}
  M.~Eingorn, J.~Nov\'ak and A.~Zhuk,
  Eur.\ Phys.\ J.\ C {\bf 74} (2014) no.8,  3005

\bibitem{Tsujikawa:2007tg}
  S.~Tsujikawa, K.~Uddin and R.~Tavakol,
  Phys.\ Rev.\ D {\bf 77} (2008) 043007

\bibitem{Tsujikawa:2007gd}
  S.~Tsujikawa,
  Phys.\ Rev.\ D {\bf 76} (2007) 023514
  doi:10.1103/PhysRevD.76.023514
  [arXiv:0705.1032 [astro-ph]].

\bibitem{Stelle:1976gc}
  K.~S.~Stelle,
  Phys.\ Rev.\ D {\bf 16} (1977) 953.
  doi:10.1103/PhysRevD.16.953

\bibitem{Mannheim:1988dj}
  P.~D.~Mannheim and D.~Kazanas,
  Astrophys.\ J.\  {\bf 342} (1989) 635.

\bibitem{Capozziello:2012ie}
  S.~Capozziello and M.~De Laurentis,
  Annalen Phys.\  {\bf 524} (2012) 545.

\bibitem{Khoury:2003rn}
  J.~Khoury and A.~Weltman,
  Phys.\ Rev.\ D {\bf 69} (2004) 044026
  
 \bibitem{CapTsu}
 S. Capozziello, and  S. Tsujikawa,
Phys. Rev. D77 (2008) 107501


\bibitem{Dialektopoulos:2017pgo}
  K.~F.~Dialektopoulos, A.~Nathanail and A.~G.~Tzikas,
  arXiv:1712.10177 [gr-qc].

\bibitem{cardone}
S. Capozziello, V. F. Cardone, V. Salzano, Phys. Rev. D, {\bf 78}, (2008), 063504.

\bibitem{altroausiliario}
S. Capozziello, V. F. Cardone, A. Troisi, Phys. Rev. D, {\bf 71}, (2005), 043503.

\bibitem{pad1}
C. Gruber, O. Luongo, Phys. Rev. D, {\bf 89},  (2014), 103506.

\bibitem{pad2}
A. Aviles, A. Bravetti, S. Capozziello, O. Luongo, Phys. Rev. D, {\bf 90}, (2014), 043531.

\bibitem{fis}
A. Aviles, C. Gruber, O. Luongo, H. Quevedo, Phys. Rev. D, {\bf 86}, (2012), 123516.

\bibitem{ioepeter}
P. K. S. Dunsby, O. Luongo, Int. J. Geom. Meth. Mod. Phys., {\bf 13}, 03, (2016), 1630002.

\bibitem{salzano}
S. Capozziello, M. Faizal, M. Hameeda, B. Pourhassan, V. Salzano, S. Upadhyay 
Mon. Not. Roy. Astron. Soc. 474 (2018) 2430.



\end{thebibliography}
\end{document}